\documentclass[12pt]{article}
\usepackage{epsfig}
\def\be{\begin{equation}}
\def\ee{\end{equation}}
\def\ba{\begin{eqnarray}}
\def\ea{\end{eqnarray}}
\def\ga{\mathrel{\raise.3ex\hbox{$>$\kern-.75em\lower1ex\hbox{$\sim$}}}}
\def\la{\mathrel{\raise.3ex\hbox{$<$\kern-.75em\lower1ex\hbox{$\sim$}}}}

\begin{document}
\baselineskip=16pt
\begin{titlepage}
\begin{center}

\vspace{0.5cm}
{\Large \bf Assisted Tachyonic Inflation }\\
\vspace{10mm}
Yun-Song Piao$^{a,c}$, Rong-Gen Cai $^{c}$, Xinmin Zhang$^{a}$ and
Yuan-Zhong Zhang$^{b,c}$ \\
\vspace{6mm}
{\footnotesize{\it
 $^a$Institute of High Energy Physics, Chinese
     Academy of Sciences, P.O. Box 918(4), Beijing 100039, China\\
 $^b$CCAST (World Lab.), P.O. Box 8730, Beijing 100080\\
 $^c$Institute of Theoretical Physics, Chinese Academy of Sciences,
      P.O. Box 2735, Beijing 100080, China \footnote{Mailing address in
      China, Email address: yspiao@itp.ac.cn}\\}}

\vspace*{5mm}
\normalsize
\smallskip
\medskip
\smallskip
\end{center}
\vskip0.6in
\centerline{\large\bf Abstract}
 {The model of inflation with a single tachyon field generates larger
anisotropy and has difficulties in describing the formation of the
Universe \cite{LK}. In this paper we consider a model with multi tachyon
fields
and study the assisted inflationary solution. Our results show
that this model satisfies the observations.
 }
\vspace*{2mm}
\end{titlepage}

%\section{Introduction}

Recently Sen\cite{S} has constructed a classical time-dependent
solution which describes the decay process of an unstable D-brane
in the open string theory. During the decay process the tachyonic
field on the brane rolls down toward the minimum of the potential.
There have been recently a lot of studies on various cosmological
effects of the rolling tachyon \cite{GMP}. In general the
unconventional form of the tachyonic action makes the cosmology
with tachyonic field differ from that with a normal scalar field.
Regarding the role of tachyon as a inflaton\cite{FC,CGJP},
recently Kofman and Linde \cite{LK} pointed out the difficulties
in generating an acceptable metric perturbation. In this paper,
we propose a model of inflation with multi-tachyon, and study the
dynamics of this type of assisted inflation. Our results show that
some of the difficulties with a single tachyonic field are overcome and
our
model satisfies the observations.

We start with a brief
review on the problems of the single tachyonic inflation model.
The 4D effective field theory of the tachyon field coupled to
Einstein gravity is given by
\be
S={1\over 2\kappa^2}\int d^4 x \sqrt{-g} R +S_{brane} ,
\label{S}
\ee
where $\kappa^2 ={1\over M_p^2}$ is the 4D Planck mass, $S_{brane}$ the
4D
effective field theory action of the tachyon field.
To a non-BPS D3-brane
action in supersymmetric theory (BSFT)\cite{KMM}, one has
\be
S_{brane}=\tau_3 \int d^4 x
\left(\alpha^\prime \ln{2} e^{-{T^2\over 4}}\partial_{\mu}
T\partial^{\mu}T + e^{-{T^2\over 4}}\right) ,
\ee
where the tension of the non-BPS brane is
\be
\tau_3 ={\sqrt{2} M_s^4\over (2\pi)^3 g_s} ,
\label{tau3}
\ee
with
$g_s$ being the string coupling, and $M_s = l^{-1}_s ={1\over
\sqrt{\alpha^\prime}}$
the fundamental string mass and length scales. The Planck
mass in this theory is given by a dimensional reduction \cite{JST}
\be
M_p^2 ={v M_s^2\over g_s^2} .
\ee
In Eq.(4),
$v={(M_s r)^d\over \pi}$, $r$ is the radius of the compactification, and
$d$ is the number of the compactified dimensions.

Redefining the tachyon field $T\rightarrow \sqrt{2\ln{2}}T$, the
effective action of the tachyon field in the Born-Infeld form can
be written as \cite{S, G}\footnote{For a scalar field $\phi $ with
standard kinetic term, the authors of \cite{SM} have given a
constraint on the potential $U$ of the scalar field: $2 H^2 <
\kappa^2 U < 3H^2$. With this potential satisfying the constraint,
the cosmological inflation can occur. It is interesting to see
whether there is a similar constraint on the potential $V$ of the
tachyon in the action (\ref{eq5}). A similar calculation yields a
constraint: $0<\kappa^2 V <3\sqrt{3} H^2$.}
 \be
\label{eq5}
  S_{brane} = \int d^4 x
\sqrt{-g} V(T)\sqrt{1+\alpha^\prime \partial_{\mu} T
\partial^{\mu} T},
\label{Sb} \ee which is a close expression incorporating all
higher order of powers of $\partial_{\mu} T$. The potential $V(T)$
around the maximum is \be V(T) \simeq \tau_3 e^{-{T^2\over 8ln2}}
, \label{vt} \ee which has a maximum at $T=0$ and at large $T$ in
terms of Ref. \cite{S}, the potential should be exponential \be
V(T)\simeq e^{-T} . \label{lvt} \ee The potential $V(T)$ in the
Born-Infeld type of action can be regarded as a smooth function
which interpolates between two asymptotic expressions given by
(\ref{vt}) at maximum and by (\ref{lvt}) at infinity.

Assuming that inflation occurs near the top of the tachyon
potential, for a tachyonic field which is spatially homogeneous
but time-dependent the slow-rolling conditions are given
by\cite{CGJP} \be \epsilon = {M_p^2\over 2\alpha^\prime V(T)}
\left({V^\prime(T)\over V(T)}\right)^2  < 1 ; \label{eps} \ee \be
\eta = {M_p^2\over \alpha^\prime V(T)}{V^{\prime\prime}\over V(T)}
< 1 . \label{eta} \ee In the slow-rolling approximation, the
Hubble expansion rate can be expressed as \be H^2 \simeq
{\kappa^2\over 3} V(T), \label{h12} \ee and the equation of motion
for the rolling tachyon in an expanding universe is \be 3H{\dot
{T}} + (\alpha^\prime V(T))^{-1} V^{\prime} \simeq 0  . \label{t1}
\ee Substituting (\ref{tau3}) and (\ref{vt}) into (\ref{eta}), we
have \be g_s > {(2\pi)^3 v\over 4\sqrt{2}\ln{2}} \sim 10^2 v .
\ee The amplitude of the gravitational waves produced during
inflation is \cite{SS} \be {\cal {P}}_G \sim {H\over M_p}\leq
3.6\times 10^{-5} . \label{pg} \ee Substituting (4) and
(\ref{h12}) into (\ref{pg}), we have \be g_s^3\leq 10^{-7} v^2 .
\ee Combining (12) and (14), one can see that for the inflation
driven by the rolling of the tachyon to happen, the following
condition \be v < 10^{-13} \label{v} \ee must be satisfied. Since
the 4D effective theory is applicable only if $v \gg 1$
\cite{JST}, Eq.(15) implies that the single tachyonic inflation
can not result in a reasonable observed universe.

To solve this problem, we
introduce more tachyon fields into the model and consider an assisted
inflation scenario with multi-tachyonic fields.
Specifically we consider a system with $n$ non-coincident but
parallel non-BPS D3-branes \cite{MPP}. In this system, there are
two kinds of open strings. One of them starts from and ends on the
same brane; The other starts from a given brane, then ends on a
different brane. If the distance between two branes are much
larger than the string scale, one can ignore the second kind of
open strings, leaving a tachyon on the world volume for every
brane. Thus, we have $n$ tachyons without interaction. And the
action is simply the sum of $n$ single-tachyonic actions:
 \be S_{brane}=\sum_{i=1}^n \int d^4 x \sqrt{-g}
V(T_i)\sqrt{1+\alpha^\prime
\partial_{\mu} T_i \partial^{\mu} T_i}. \ee

With the multi-tachyonic fields the slow-rolling conditions are
found to be
 \be \epsilon = {M_p^2\over 2\alpha^\prime V_{sum}(T)}
\left({V^\prime(T_i)\over V(T_i)}\right)^2  < 1 ; \label{neps}
\ee
 \be \eta = {M_p^2\over \alpha^\prime
V(T_i)}{V^{\prime\prime}(T_i)\over V_{sum}(T)} < 1 , \label{neta}
\ee
 where $V_{sum}(T)=\sum_{i=1}^n V(T_i)$. The Friedman equation
is now
  \be H^2 \simeq {\kappa^2\over 3}V_{sum}(T) . \label{h2}
   \ee
 And the equation of motion for one of the tachyons is
  \be 3H{\dot{T}}_i + (\alpha^\prime V(T_i))^{-1} V^{\prime}(T_i)
   \simeq 0 . \label{t}
 \ee
  Similar to the discussions on the constraints on the
model parameters above,  for the assisted inflation to be
successful we have
 \be n g_s^3\leq 10^{-7} v^2 ;~~~~n g_s > 10^2 v .
 \ee
 Thus we get,
  \be v < 10^{-13} n^2 . \ee
   One can see that
when $n\geq 10^{7}$, the condition, $v\gg 1$, required by the
applicability of the 4D effective field theory is satisfied.

In the following, we will calculate the amplitude of the density
perturbation of our model. For potential (\ref{vt}) and the
equations of motion of the tachyon fields (\ref{t}), we notice that
the equation of motion for each of the tachyon fields follow a
simple relationship
 \be {d \ln{T_i}\over dt}\simeq {d\ln{T_1}\over dt}. \ee
  This means that if the slow-rolling conditions are satisfied,
all the tachyons would follow a similar trajectory with a unique
late attractor, {\it i.e.} $T_1\sim T_2 ...\sim T_n\equiv T$. The
calculation of the density perturbation responsible for the
anisotropy of CMB depends crucially on this late-time attractor of
the fields. In this case, the Friedman equation (\ref{h2}) can be
rewritten as
 \be H^2 \simeq {\kappa^2\over 3} n V(T) , \ee
 where $V(T)$ at maximum is given by (\ref{vt}). The number of e-folds
during inflation is
 \be N=\int H dt \simeq
-\int_{T_{60}}^{T_{end}}{H^2 V(T)\over V^\prime (T)} dT ,
\label{n} \ee
 where $T_{60}$ is the field value corresponding to
$N\simeq 60$ as required when the COBE scale exits the Hubble
radius, and $T_{end}$ is the value of the tachyon field at which inflation
ends.
And $T_{end}$ is determined by
 \be \eta \simeq {10^2 v\over n g_s}
e^{{T_{end}^2\over 8\ln{2}}}\simeq 1 . \label{eta1} \ee
 Thus, we have
  \be T_{end}^2 \simeq 8\ln{2} \ln{{n g_s\over 10^2 v}} .
\label{tend} \ee

Following the definition of the amplitudes of the density
perturbation in Refs. \cite{SS} and \cite{LMS}, we have
 \be {\cal
{P}}_S\sim {1\over \sqrt{\alpha^\prime n V(T)}}{H^2\over 2\pi
{\dot {T}}} , \label{ps} \ee
 and
  \be {\cal {P}}_S\sim {n g_s^{2}\over v^{{3\over 2}}T_{60}}.
   \label{ps1} \ee
 Taking $n\sim 10^{11}$, $v\sim 10$ and $g_s\sim 10^{-8}$,
 from (\ref{n}) and
(\ref{tend}), we obtain that $10 T_{60}\sim T_{end}\sim {\cal
{O}}(1)$. Substituting the values of these parameters into
(\ref{ps1}), we have the amplitudes of the density perturbation,
${\cal {P}}_S \sim 10^{-6}$. This is consistent with the
observation of COBE. In this case, from (4) and (\ref{tau3}), the
string mass scale $M_s\sim 10^{-8} M_p$ and the total tension
$n\tau_3 \sim 10^{-16} M_p^4$.

%\vspace{100mm}

In summary we have proposed a multi-tachyonic inflation model. Our
results shown that with the help of a large number of tachyons the
problems with  a single tachyonic field to satisfy the
slow-rolling condition and to generate the amplitude of the
gravitational waves constrained by the CMB anisotropy can be
solved. Furthermore the energy scale associated with the inflation
in our model is also much smaller than the Planck scale. We should
point out, however that our model does not solve all of the
problems raised in \cite{LK}. In fact $M_s \leq H$ required in the
single as well as multi tachyonic inflation model implies that the
size of the de-Sitter horizon will be smaller than the string
length {\it i.e.} ${1\over H} \leq l_s$,  which makes it invalid
to describe the tachyon condensation by using an effective field
theory \cite{LK}
 \footnote{We thank A. Linde for pointing out this to
us. }. In the general inflation models, the inflaton mass is of
the same order as or less than $H$. However in the tachyonic
inflation model the inflaton has a mass around the string scale
$M_s =l_s^{-1}$. There might be some possible ways to overcome
this difficulty. For instance, one may extend the model studied in
this paper by including the effects of tachyons resulting from the
open string stretching between different branes. There could exist
the possibility in such kind of multi-brane system that the
inflaton (tachyon) mass scale be fine-tunned so that it is in the
range required and smaller than the string scale $M_s$. To realize
this speculation with a model may not be easy and goes beyond the
scope of this paper. Finally, we would also like to mention here
that the usual reheating mechanism is problematic in the tachyon
inflation model, as pointed out by Kofman and Linde \cite{LK},
because the tachyon does not oscillate during the decay of non-BPS
branes. This issue has been discussed recently in \cite{STW}.

This work implies that the inflation and
cosmological applications of the multi-tachyon/multi-brane
may have more fruitful phenomena, which is worth
studying further.

\textbf{Acknowledgments}

We would like to thank Robert Brandenberger, Xiaobo Gong, Qing-Guo
Huang, Miao Li and Dehai Zhang for
discussions. This project was in part supported by
NNSFC under Grant Nos. 10175070 and 10047004, as well as also by the
Ministry of Science and Technology of China under grant No.
NKBRSF G19990754.

\end{document}